\documentclass{PoS}

\usepackage{axodraw}

\newcommand{\be}{\begin{equation}}
\newcommand{\ee}{\end{equation}}
\newcommand{\ba}{\begin{eqnarray}}
\newcommand{\ea}{\end{eqnarray}}

\title{Chiral perturbation theory in the meson sector}

\ShortTitle{ChPT in the meson sector}

\author{\speaker{Johan Bijnens}\\%
        Department of Theoretical Physics, Lund University,
        S\"olvegatan 14A, SE 22362 Lund, Sweden\\
        E-mail: \email{bijnens@thep.lu.se}}

\abstract{The present status of Chiral Perturbation Theory in the meson sector
is discussed concentrating on recent developments. 
This write-up contains short discussions
on a listing a few historical papers,
the principles behind ChPT,
two-flavour ChPT including some comments about the  pion polarizability,
three-flavour ChPT with a discussion of the recently found relations
as tests of ChPT and
preliminary results of a new fit of the NLO low-energy-constants (LECs).
It discusses somewhat deeper $\eta\to3\pi$ and the arguments for the
existence of a ``hard pion ChPT'' and its application to $K\to2\pi$.
}

\FullConference{6th International Workshop on Chiral Dynamics\\
                 July 6-10 2009\\
                 Bern, Switzerland}

\renewcommand{\logo}
{\raisebox{-5mm}
{\parbox{\textwidth}{\begin{flushright}
LU TP 09-25\\
September 2009
\end{flushright}
\includegraphics{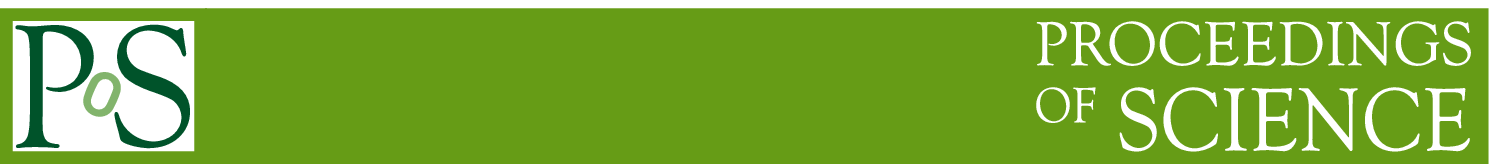}}
}}
\FullConference{{\rm To be published in the proceedings of:\\}
6th International Workshop on Chiral Dynamics\\
                 July 6-10 2009\\
                 Bern, Switzerland}
\begin{document}

\section{Introduction}

Chiral Perturbation Theory (ChPT) and effective field theory (EFT)
play a major role at this now 15 year old conference series. 
In this talk I review some aspects of mesonic ChPT.
This talk has a large overlap with earlier talks given at Lattice 2007
\cite{lattice07} and EFT09 \cite{valencia09} as well as with my review
on ChPT at two-loop order \cite{reviewp6}. I concentrate on some of the newer
results in this written version, consult the earlier references for
the topics discussed here in less detail.

In the talk I discussed several topics not included
in this write-up.
The discussion and references for the partially quenched calculations
and recent progress in renormalization group and effective field
theory can be found
in \cite{valencia09} and for the 
results for ChPT for the weak interaction
a recent review is \cite{reviewK}.

This write-up contains short discussions
on a listing a few historical papers,
the principles behind ChPT,
two-flavour ChPT including some comments about the  pion polarizability,
three-flavour ChPT with a discussion of the recently found relations and
preliminary results of a new fit of the NLO low-energy-constants (LECs).
I devote quite some space to $\eta\to3\pi$ and the arguments for the
existence of a ``hard pion ChPT'' and its application to $K\to2\pi$.

\section{Some History: 50, 40, 35, 30, 25, 20 and 15 years ago}
\label{history}

In this ``capital city'' of Chiral Perturbation Theory
it is appropriate to look back at some of its history.
ChPT has a 50 year history by
now as was reviewed by S.~Weinberg in his talk \cite{Weinbergfirst}.
It should also be remarked that this conference series is now 15 years old.

I picked some papers
which fell at or close to jubileum years. About 50 years ago our subject
started with the Goldberger-Treiman relation \cite{GB} and
the advent of PCAC, the partially conserved axial-current \cite{PCAC},
and how this reproduced the Goldberger-Treiman relation.
About 40 years ago a lot of work had been done within the framework
of PCAC but 1968 and 1969 saw some very important papers:
the
Gell-Mann--Oakes--Renner relation \cite{GMOR} and the proper way how to
implement chiral symmetry in all generality in phenomenological
Lagrangians \cite{CCWZ} after Weinberg's derivation for the two flavour case
\cite{Weinberg1}. Shortly afterwards loop calculations started
with e.g. loop results for $\pi\pi$ scattering \cite{EH}
and $\eta\to3\pi$ \cite{LP}. 30 years ago the start with the modern way of
including higher order Lagrangians and performing a consistent renormalization
came with \cite{Weinberg0}. At the same time there was also the beautiful
paper by Gasser and Zepeda about the types of non-analytical corrections that
can appear \cite{GZ}. The seminal papers by Gasser and Leutwyler of 25 years
ago then put the entire subject on a modern firm footing \cite{GL0,GL1}.
The same period also had my own entry into the subject \cite{BSW}.
Many one-loop calculations were done and the understanding that
the coefficients in the higher-order Lagrangians could be understood
from the contributions of resonances
was put on a firm footing 20 years ago \cite{GL0,EGPR}.
Let me close this historical part with two 15 year old papers,
a very clear discussion of the basics of ChPT \cite{Leutwyler1} and
the first full two-loop calculation \cite{BGS}.

\section{Chiral Perturbation Theory: ChPT, CHPT or $\chi$PT}

ChPT is best described as
``\emph{
Exploring the consequences of the chiral symmetry of QCD
and its spontaneous breaking using
effective field theory techniques}''
and a clear discussion about its derivation and underlying
assumptions is in \cite{Leutwyler1}.
Some reviews are
\cite{review1,reviewp6}. More reviews and
references to introductory lectures can be found on the webpage
\cite{website}. 

For effective field theories, there are three principles that are needed
and for ChPT they are
\begin{itemize}
\parskip0cm\itemsep0cm
\item {\bf Degrees of freedom:} Goldstone Bosons from
the spontaneous chiral symmetry breakdown.
\item {\bf Power counting:} This is what allows a systematic ordering of terms
and is here essentially dimensional counting in momenta and masses.
\item {\bf Expected breakdown scale:} The scale of the not explicitly included
physics, here resonances, so the scale is of order $M_\rho$, 
but this is channel dependent. 
\end{itemize}
I will not go into more details here, links to lectures can be found
on the website \cite{website}, short introductions can be found
in \cite{valencia09} and \cite{reviewp6}.

\section{Two-flavour ChPT at NNLO}

References to order $p^2$ and $p^4$ work can be found in \cite{reviewp6}.
The first work at NNLO used dispersive methods to obtain the nonanalytic
dependence on kinematical quantities, $q^2,s,t,u$ at NNLO. This was done
for the vector (electromagnetic) and scalar form-factor of the pion
in \cite{GM} (numerically) and \cite{CFU} (analytically) and for
$\pi\pi$-scattering analytically in \cite{Knechtpipi}. 
%The work of
%\cite{Knechtpipi} allowed to put many full NNLO calculations
%in two-flavour ChPT in a simple analytical form.

Basically all processes of interest are calculated to NNLO in ChPT:
$\gamma\gamma\to\pi^0\pi^0$ \cite{BGS,GIS1},
$\gamma\gamma\to\pi^+\pi^-$ \cite{Burgi1,GIS2},
$F_\pi$ and $m_\pi$ \cite{Burgi1,BCEGS1,BCT}, $\pi\pi$-scattering
\cite{BCEGS1}, the pion scalar and vector form-factors \cite{BCT}
and pion radiative decay $\pi\to\ell\nu\gamma$ \cite{BT1}.
The pion mass is known at order $p^6$ in finite volume \cite{CH}.
Recently $\pi^0\to\gamma\gamma$ has been done to this order
as discussed in the talk by Moussallam\cite{Kampf1}.

The LECs have been fitted in several processes. $\bar l_4$ from fitting
to the pion scalar radius \cite{BT1,CGL}, $\bar l_3$ from an estimate of
the pion mass dependence on the quark masses \cite{GL0,CGL}
and $\bar l_1$, $\bar l_2$ from the agreement with 
$\pi\pi$-scattering \cite{CGL}, $\bar l_6$ from the pion charge radius 
\cite{BCT}
and $\bar l_6-\bar l_5$ from the axial form-factor in $\pi\to\ell\nu\gamma$.
There is also a recent determination of $\bar l_5$ from hadronic
tau decays \cite{PP}.
The final best values are \cite{BCT,BT1,CGL,PP}
\be
\label{valueli}
\begin{array}{llll}
\bar l_1=-0.4\pm 0.6\,,\quad&
\bar l_2 =4.3\pm0.1\,,&
\bar l_3=2.9\pm2.4\,,&
\bar l_4=4.4\pm0.2\,,
\\
\bar l_6-\bar l_5 = 3.0\pm0.3\,,\quad&
\bar l_6 = 16.0\pm0.5\pm0.7\,,\quad&
\bar l_5 = 12.24\pm0.21\,.&
\end{array}
\ee
Values of $\bar l_3$ and $\bar l_4$ have also been obtained by the lattice as
discussed in several talks at this conference.
There is information on some combinations of $p^6$ LECs.
These are basically via the curvature in the vector and scalar form-factor
of the pion \cite{BCT} and two combinations from $\pi\pi$-scattering
\cite{CGL} from the knowledge of $b_5$ and $b_6$ in that reference.
The order $p^6$ LECs $c_i^r$ are estimated 
to have a small effect for $m_\pi,f_\pi$ and $\pi\pi$-scattering.

A possible problem for ChPT are the pion polarizabilities.
These are cleanly predicted in ChPT, the latest numbers
from ChPT \cite{GIS2} and experiment  \cite{Mainz2005} are:
\ba
\mathrm{ChPT:~} 
(\alpha_1-\beta_1)_{\pi^\pm} &=& (5.7\pm1.0)\cdot 10^{-4}~\mathrm{fm}^3
\\
\label{pollarge}
\mathrm{Exp:~~}
(\alpha_1-\beta_1)_{\pi^\pm} &=& (11.6\pm1.5_{stat}\pm3.0_{syst}\pm0.5_{mod})
\cdot 10^{-4}~\mathrm{fm}^3
\ea
A possible problem in this experiment
is the background from direct $\gamma N\to \gamma N \pi$
production. Large values also follow from the older
Primakoff experiments and a dispersive analysis \cite{Filkov}
from $\gamma\gamma\to\pi\pi$:
\ba 
\mathrm{Primakoff:~} (\alpha_1-\beta_1)_{\pi^\pm}
& =& (13.6\pm2.8_{stat}\pm2.4_{syst})
\cdot 10^{-4}~\mathrm{fm}^3
\\
\mathrm{dispersive:~}
(\alpha_1-\beta_1)&=&(13.0+3.6-1.9)\cdot10^{-4}\mathrm{fm}^3\,.
\ea
The latter value has been criticized in \cite{Drechsel}
who argue that \cite{Filkov} has a large uncontrolled
model dependence and conclude ``
``Our calculations\ldots are in 
reasonable agreement with ChPT for charged pions''.
See their presentations in these proceedings for more details.
ChPT in our present understanding cannot produce
 a value as large as in (\ref{pollarge}).

\section{Three-flavour ChPT}

\subsection{Calculations}

In this section I  discuss several results at NNLO in mesonic three-flavour
ChPT. The formulas here are much larger than in
two-flavour ChPT and while the expressions have been reduced to a series
of well-defined two-loop integrals, the latter are evaluated numerically.
Both are the consequence of the different masses present.
The vector two-point functions
\cite{ABT1,GK1} and the isospin breaking in the $\rho\omega$
channel \cite{Maltman} were among the first calculated.
The flavour disconnected scalar two-point function
relevant for bounds on $L_4^r$ and $L_6^r$ was worked out in \cite{Moussallam}.
The remaining scalar two-point functions are known,
available from the speaker but unpublished.
Masses and decay constants as well as axial-vector
two-point functions were the first calculations requiring full two-loop
integrals, done in the $\pi$ and $\eta$ \cite{ABT1,GK2} and the $K$ channel
\cite{ABT1}. Including isospin breaking contributions to masses and decay
constants was done in \cite{ABT4}.
After $K_{\ell4}$ had also been
evaluated to NNLO \cite{ABT3} a fit to the LECs was done as described
below. The vacuum expectation values
in the isospin limit were done in \cite{ABT3},
with isospin breaking in \cite{ABT4} and at finite volume in~\cite{BG1}.

Vector (electromagnetic) form-factors for pions and kaons were calculated
in \cite{PS1,BT2} and in \cite{BT2} a NNLO fit for $L_9^r$ was performed.
$L_{10}^r$ can be had from hadronic tau decays \cite{PP}
or the axial form-factor in $\pi,K\to\ell\nu\gamma$.
The NNLO calculation is done, but no data fitting was performed \cite{Geng}.
A rather important calculation is the $K_{\ell3}$ form-factor. This calculation
was done in \cite{BT3,PS3} and a rather interesting relation between
the value at zero, the slope and the curvature for the scalar form-factor
obtained \cite{BT3}. Isospin-breaking has been included as well \cite{BG3}.

Scalar form-factors including sigma terms and scalar radii
\cite{BD} and $\pi\pi$ \cite{BDT1} and $\pi K$-scattering \cite{BDT2}
are known and used to place limits on $L_4^r$ and $L_6^r$.
Finally, the relations between the $l_i^r,c_i^r$ and $L_i^r,C_i^r$
have been extended
to the accuracy needed to compare order $p^6$ results in two and three-flavour
calculations \cite{Haefeli2} and there has been some progress towards fully
analytical results for
$m_\pi^2$ \cite{Kaiser} and $\pi K$-scattering lengths \cite{KS}.
The most recent results are
 $\eta\to3\pi$ \cite{BG2}, isospin breaking in $K_{\ell3}$ \cite{BG3}.

\subsection{Testing ChPT and estimates of the order $p^6$ LECs}

Most numerical analysis at order $p^6$ use a (single) resonance
approximation to the order $p^6$ LECs. 
The main underlying motivation is the large $N_c$ limit and phenomenological
success at order $p^4$ \cite{EGPR}.
There is a large volume of work on this,
some references are \cite{MHA1}.
The numerical work I will report has used a rather simple resonance Lagrangian
\cite{EGPR,BCEGS1,ABT4,ABT3}.
%\ba
%\label{lagreso}
%{\cal L}_V &=& -\frac{1}{4}\langle V_{\mu\nu}V^{\mu\nu}\rangle
%+\frac{1}{2}m_V^2\langle V_\mu V^\mu\rangle
%-\frac{f_V}{2\sqrt{2}}\langle V_{\mu\nu}f_+^{\mu\nu}\rangle
%-\frac{ig_V}{2\sqrt{2}}\langle V_{\mu\nu}[u^\mu,u^\nu]\rangle
%+f_\chi\langle V_\mu[u^\mu,\chi_-]\rangle\,,
%\nonumber\\
%{\cal L}_A &=& -\frac{1}{4}\langle A_{\mu\nu}A^{\mu\nu}\rangle
%+\frac{1}{2}m_A^2\langle A_\mu A^\mu\rangle
%-\frac{f_A}{2\sqrt{2}}\langle A_{\mu\nu}f_-^{\mu\nu}\rangle\,,
%\nonumber\\
%{\cal L}_S &=& \frac{1}{2} \langle \nabla^\mu S \nabla_\mu S 
% - M^2_S S^2 \rangle  
% + c_d \langle Su^\mu u_\mu \rangle + c_m \langle S \chi_+ \rangle\,, 
%\,\,\,
%{\cal L}_{\eta^\prime} = \frac{1}{2}\partial_\mu P_1\partial^\mu P_1
%-\frac{1}{2}M_{\eta^\prime}^2 P_1^2
%+i\tilde d_m P_1\langle\chi_- \rangle\,.
%\nonumber\\
%f_V &=& 0.20,\quad  f_\chi = -0.025,\quad  g_V = 0.09, 
%\quad  c_m = 42 \mbox{ MeV},\quad
%  c_d = 32 \mbox{ MeV}, \quad
% \tilde d_m = 20 \mbox{ MeV}, 
%\nonumber\\
%m_V &=& m_\rho = 0.77 \mbox{ GeV},  m_A = m_{a_1} = 1.23 \mbox{ GeV}, 
%m_S = 0.98 \mbox{ GeV},\quad m_{P_1} =  0.958 \mbox{ GeV}\,.
%\ea
%The values of $f_V$, $g_V$, $f_\chi$ and $f_A$ come from experiment
% \cite{BCEGS2,Ecker1} and
%$c_m$ and $c_d$ from resonance saturation at order $p^4$ \cite{Ecker1}. 
The estimates of the $C_i^r$ is the weakest point in the numerical
fitting at present, however, many results are not very sensitive to this.
The main problem is that the $C_i^r$ which contribute to the masses,
are estimated to be zero, except for $\eta'$ effects, and how these
might affect the determination of the others. 
The estimate is $\mu$-independent
while the $C_i^r$ are not.

The fits done in \cite{ABT4,ABT3,BD} tried to check this by varying
the total resonance contribution by a factor of two, varying the scale $\mu$
from $550$ to $1000$~MeV and compare estimated $C_i^r$ to experimentally
determined ones. The latter works well, but the experimentally 
well determined ones are those with dependence on kinematic variables only,
not ones relevant for quark-mass dependence.

A new fit is in progress but in order to check whether ChPT with
three flavours works, one would like a test that is as much as possible
independent of the estimated values of the $C_i$. In \cite{BJ}
we studied 76 observables leading to 35 combinations that are independent
of the $C_i$, or 35 relations. For these we found 13 with good ``data,''
$\pi\pi$ \cite{CGL} and $\pi K$ \cite{BDM}threshold parameters from
dispersion theory
and $K_{\ell4}$ from experiment \cite{Pislak1,NA48}.

The results for the 5 relations found in $\pi K$ scattering are shown
in Tab.~\ref{tab:piK}. The equality of the remainder of LHS and RHS gives
a test of ChPT. The results are encouraging but not 100\% conclusive.
More details and more results can be found in \cite{BJ,BJtalk}.
\begin{table}
\begin{center}
\begin{tabular}{|c|c|r|r|r|r|c|}
\hline
                  & Roy-Steiner        & NLO   & NLO  &NNLO    &NNLO    & remainder\\
                  &                    & 1-loop&LECs  &2-loop  &1-loop  &          \\
\hline
LHS (I)       & $ 5.4\pm0.3  $&$0.16$&$ 0.97$&$ 0.77$&$-0.11$&$ 0.6\pm0.3$\\
RHS (I)       & $ 6.9\pm0.6  $&$0.42$&$ 0.97$&$ 0.77$&$-0.03$&$ 1.8\pm0.6$\\
\hline
10 LHS (II)    & $ 0.32\pm0.01$&$0.03$&$ 0.12$&$ 0.11$&$ 0.00$&$ 0.07\pm0.01$\\
10 RHS (II)    & $ 0.37\pm0.01$&$0.02$&$ 0.12$&$ 0.10$&$-0.01$&$ 0.14\pm0.01$\\
\hline
100 LHS (III)   & $-0.49\pm0.02$&$0.08$&$-0.25$&$-0.17$&$ 0.05$&$-0.21\pm0.02$\\
100 RHS (III)   & $-0.85\pm0.60$&$0.03$&$-0.25$&$ 0.11$&$-0.03$&$-0.71\pm0.60$\\
\hline
100 LHS (IV)   & $ 0.13\pm0.01$&$0.04$&$ 0.00$&$ 0.01$&$ 0.03$&$ 0.05\pm0.01$\\
100 RHS (IV)   & $ 0.01\pm0.01$&$0.01$&$ 0.00$&$ 0.00$&$ 0.00$&$-0.01\pm0.01$\\
\hline
$10^3$ LHS (V)& $ 0.29\pm0.05$&$0.09$&$ 0.00$&$ 0.06$&$ 0.01$&$ 0.13\pm0.03$\\
$10^3$ RHS (V)& $ 0.31\pm0.07$&$0.03$&$ 0.00$&$ 0.06$&$ 0.05$&$ 0.17\pm0.07$\\
\hline
\end{tabular}
\end{center}
\caption{\label{tab:piK}
$\pi K$-scattering: The numerical results for relations I-V,
both left- (LHS) and right-hand side for the dispersive
result from \cite{BDM} and the NLO, NNLO 2-loop and NNLO $L_i$-dependent
part (1-loop). The tree level for LHS and RHS
 of (I) is 3.01 and vanishes for the others.
The equality of the remainder of LHS and RHS gives a test of ChPT.
}
\end{table}

\subsection{The fitting and results}

The inputs used for the standard fit, as discussed more extensively in
\cite{ABT4,ABT3}, are
\begin{itemize}
\parskip0cm\itemsep0cm
\item $K_{\ell4}$: $F(0)$, $G(0)$, $\lambda$ from E865 at 
BNL\cite{Pislak1}.
\item $m^2_{\pi^0}$, $m^2_\eta$, $m_{K^+}^2$, $m_{K^0}^2$, electromagnetic
corrections include the violation of Dashen's theorem.
%(\cite{Dashen} and references therein). 
\item $F_{\pi^+}$ and $F_{K^+}/F_{\pi^+}$.
\item
$m_s/\hat m = 24$. Variations with
$m_s/\hat m$ were studied in \cite{ABT4,ABT3}.
\item
$L_4^r, L_6^r$ the main fit, 10, has them equal to zero, but see below
and the arguments in \cite{Moussallam}.
\end{itemize}
Some results of this fit are given in Tab.~\ref{tabfits}.
The errors are very correlated, see Fig.~6 in \cite{ABT3} for an example.
Varying the values of $L_4^r,L_6^r$ as input can be done with a
reasonable fitting chi-squared when varying $10^3 L_4^r$ from $-0.4$ to $0.6$
and $L_6^r$ from $-0.3$ to $0.6$ \cite{BD}.
The variation of many quantities with $L_4^r,L_6^r$
(including the changes via the changed values of the other $L_i^r$) are shown
in \cite{BD,BDT1,BDT2}. Fit B was one of the fits with a good fit to the
pion scalar radius and fairly small corrections to the sigma terms \cite{BD}
while fit D \cite{Kazimierz} is the one that gave agreement with
$\pi\pi$ and $\pi K$-scattering threshold quantities.

%One point should be observed here, if one fits lattice data with NLO formulas
%to obtain the $L_i^r$, one should also use the NLO or order $p^4$ fit values
%from Tab.~\ref{tabfits} to compare with. 
\begin{table}[ht]
\begin{center}
\small
\begin{tabular}{ccccc}
                & fit 10 & same $p^4$ & fit B & fit D\\
\hline
$10^3 L_1^r$ & $0.43\pm0.12$ & $0.38$ & $0.44$ & $0.44$\\
$10^3 L_2^r$ & $0.73\pm0.12$ & $1.59$ & $0.60$ & $0.69$\\
$10^3 L_3^r$ & $-2.53\pm0.37$ & $-2.91$ &$-2.31$&$-2.33$\\
$10^3 L_4^r$ & $\equiv0$    & $\equiv 0$& $\equiv0.5$ & $\equiv0.2$\\
$10^3 L_5^r$ & $0.97\pm0.11$& $1.46$ & $0.82$ & $0.88$\\
$10^3 L_6^r$ & $\equiv0$    & $\equiv 0$& $\equiv0.1$ & $\equiv0$\\
$10^3 L_7^r$ & $-0.31\pm0.14$&$-0.49$ & $-0.26$ & $-0.28$\\
$10^3 L_8^r$ & $0.60\pm0.18$ & $1.00$ & $0.50$ & $0.54$\\
$10^3 L_9^r$ & $5.93\pm0.43$ & $7.0$  & --      &  -- \\
\hline
$2 B_0 \hat m/m_\pi^2$ & 0.736 & 0.991 & 1.129 & 0.958\\
$m_\pi^2$: $p^4,p^6$    & 0.006,0.258 & 0.009,$\equiv0$ & $-$0.138,0.009 &
          $-$0.091,0.133\\
$m_K^2$: $p^4,p^6$    & 0.007,0.306 & 0.075,$\equiv0$ & $-$0.149,0.094 &
          $-$0.096,0.201\\
$m_\eta^2$: $p^4,p^6$    & $-$0.052,0.318 & 0.013,$\equiv0$ & $-$0.197,0.073 &
          $-$0.151,0.197\\
$m_u/m_d$    & 0.45$\pm$0.05 & 0.52 & 0.52 & 0.50\\
\hline
$F_0$ [MeV]          & 87.7 & 81.1 & 70.4 & 80.4 \\
$\frac{F_K}{F_\pi}$: $p^4,p^6$ & 0.169,0.051 & 0.22,$\equiv0$ & 0.153,0.067 &
   0.159,0.061
\end{tabular}
\end{center}
\normalsize
\caption{The fits of the $L_i^r$ and some results, see text for
a detailed description. They are all quoted at $\mu=0.77$~GeV.
Table with values from \protect\cite{ABT4,BT2,BD,BDT2,Kazimierz}.}
\label{tabfits}
\end{table}

Note that $m_u/m_d=0$ is never even
close to the best fit and this remains true for the entire variation
with $L_4^r,L_6^r$. The value of $F_0$,
the pion decay constant in the three-flavour chiral limit,
can vary significantly, even though I believe that fit B is an extreme case.

We, JB and I.~Jemos, are working on a new general fit.
The preferred value of $F_K/F_\pi$ is changed
and NA48 has measured the $K_{\ell4}$ formfactors more accurately.
In addition, we want to include more constraints directly.
Some preliminary results are also discussed in I.~Jemos talk \cite{BJtalk}.
For this work, all the calculated processes are being programmed
in $C++$ to allow for a more uniform treatment and an easier handling
of the LECs. This program is only partly completed but first fitting results
are given in Tab.~\ref{tab:newfit}.
The column labeled fit 10 iso uses the input as used for fit 10 in \cite{ABT4}
but without isospin breaking. The results are essentially identical
to those of the fit including isospin breaking.
This column is included as reference. The estimates for the $C_i^r$
used are the same as used in \cite{ABT4} for all the fits shown in the table
except the last.
The other columns always have some more constraints included.
The small boxes indicate the LECs which have changed most.
First we add the better information on the $K_{\ell4}$ form-factors
from NA48 \cite{NA48}. This produces sizable changes
in $L_1^r$ and $L_3^r$. The newer value of the PDG for
$F_K/F_\pi=1.193$ then changes the fitted value of $L_5^r$ which
influences $L_8^r$ via the fitted masses.
$L_5^r$ gets lowered somewhat more when we include the scattering
lengths $a^0_0$, $a^2_0$, $a^{1/2}_0$ and $a^{3/2}_0$.
Adding the pion scalar radius
requires a nonzero for one of $L_4^r$ and $L_6^r$.
This is what is shown in the column labeled ``All.''
In the last column we have set the estimated value of the $C_i^r=0$.
\begin{table}
\begin{center}
\begin{tabular}{ccccccc}
             & fit 10 iso   & NA48 & $F_K/F_\pi$ & Scatt &All         & All ($C^r_i=0$)\\
\hline
$10^3 L_1^r$ & $0.40\pm0.12$ &\framebox{$0.98$}& $0.97$&$0.97$&$0.98\pm0.11$        &$0.75$\\
$10^3 L_2^r$ & $0.76\pm0.12$ & $0.78$& $0.79$     &$0.79$&$0.59\pm0.21$        &$0.09$\\
$10^3 L_3^r$ & $-2.40\pm0.37$&\framebox{$-3.14$}&$-3.12$&$-3.14$&$-3.08\pm0.46$      &$-1.49$\\
$10^3 L_4^r$ & $\equiv0$& $\equiv0$& $\equiv0$    &$\equiv0$&\framebox{$0.71\pm0.67$}&$0.78$\\
$10^3 L_5^r$ & $0.97\pm0.11$ &$0.93$&\framebox{ $0.72$} &\framebox{$0.56$}&$0.56\pm0.11$   &$0.67$\\
$10^3 L_6^r$ & $\equiv 0$& $\equiv 0$& $\equiv 0$ &$\equiv0$&\framebox{$0.15\pm0.71$}&$0.18$\\
$10^3 L_7^r$ & $-0.30\pm0.15$&$-0.30$&$-0.26$     &$-0.23$&$-0.22\pm0.15$      &$-0.24$\\
$10^3 L_8^r$ & $0.61\pm0.20$ &$0.59$&\framebox{$0.48$}  &$0.44$&$0.38\pm0.18$        &$0.39$\\
\hline
$\chi^2$ (dof)    & $0.25$ (1) &$0.17$ (1)&$0.19$ (1)  &$5.38$ (5)&$1.44$ (4) &$1.51$ (4)
\end{tabular}
\end{center}
\caption{\label{tab:newfit} The changes in the $L_i$ compared
to the isospin symmetric fit to the same input as
the old fit 10 of \cite{ABT4}. The other columns include one new effect
at a time, see text.}
\end{table}
This is work in progress but some puzzles appear at this stage,
the large $N_c$ relation $2L_1^r = L_2^r$ is now badly broken and the
central value of $L_4^r$ is not small compared
to $L_5^r$. We are working on including more scattering lengths
and trying to include also some lattice results on the
meson masses and decay constants. This is
especially important since the simple estimate of the $C_i^r$
used has none contributing to masses and decay constants.

\section{$\eta\to \pi\pi\pi$}
\label{sectioneta}

In the limit of conserved isospin, no electromagnetism and
$m_u=m_d$, the $\eta$ is stable. Direct electromagnetic effects are 
small \cite{Sutherland1}.
The decay thus proceeds mainly through the quark-mass difference $m_u-m_d$.
The lowest order was done in \cite{orderp2x1},
order $p^4$ in
\cite{GL3} and recently the full order $p^6$ has been evaluated \cite{BG2}.
The momenta for the decay $\eta\to\pi^+\pi^-\pi^0$ we label as
$p_\eta$, $p_+$, $p_-$ and $p_0$ respectively and we introduce the
kinematical Mandelstam variables
$ %\be
s = (p_+ +p_-)^2\,, %  = (p_\eta - p_0)^2   \,,
%\nonumber\\
t = (p_+ +p_0)^2\,, %   = (p_\eta -p_-)^2     \,,
%\nonumber\\
u = (p_- +p_0)^2\,. %   = (p_\eta -p_+)^2      ,.
%\label{defstu}
$ %\ee
These are linearly dependent,
$%\bee
s+t+u = m_{\pi^{o}}^2 + m_{\pi^{-}}^2 + m_{\pi^{+}}^2 + m_{\eta}^2
\equiv 3 s_0\,.  
$%\eee
The amplitudes for the charged, $A(s,t,u)$, and
neutral, $\overline{A}(s,t,u)$ are related
\ba
\overline{A}(s_1,s_2,s_3) &=& A(s_1,s_2,s_3)+A(s_2,s_3,s_1)+A(s_3,s_1,s_2)\, .
\label{defamplitude}
\ea
The relation in (\ref{defamplitude}) is only valid
to first order in $m_u-m_d$. The overall factor of $m_u-m_d$ can be put
in different quantities, two common choices are
\ba
A(s,t,u) = \frac{\sqrt{3}}{4R}M(s,t,u)
\quad&\mbox{or}&
\quad
A(s,t,u) =\frac{1}{Q^2} \frac{m_K^2}{m_\pi^2}(m_\pi^2-m_K^2)\,
\frac{ { {\cal M}(s,t,u)}}{3\sqrt{3}F_\pi^2}\,,
\label{defM}
\ea
with %quark-mass ratios 
$R= (m_s-\hat m)/(m_d-m_u)$
or $Q^2 = R(m_s+m_d)/(2\hat m)$ pulled out.
The lowest order result is
\be
M(s,t,u)_{LO} = \left(({4}/{3})\,m_\pi^2-s\right)/F_\pi^2\,.
\label{LO}
\ee
The tree level determination of $R$ in terms of meson masses gives
with (\ref{LO}) a decay rate of 66~eV which should be compared with
the experimental results of 295$\pm$17~eV\cite{PDG06}.
In principle, since the decay rate is proportional to $1/R^2$ or $1/Q^4$,
this should allow for a precise determination of $R$ and $Q$. However,
the change required seems large. The order $p^4$ calculation
\cite{GL3} increased the predicted decay rate to 150~eV albeit with a
large error. About half of the enhancement in the amplitude came from
$\pi\pi$ rescattering and the other half from other effects like the
chiral logarithms \cite{GL3}. The rescattering effects have been
studied at higher orders using dispersive methods in \cite{KWW}
and \cite{AL}. Both calculations found an enhancement in the decay rate
to about 220~eV but differ in the way the Dalitz plot distributions
look. 
That difference and the facts that in $K_{\ell4}$ the dispersive estimate
\cite{BCG}
was about half the full ChPT calculation \cite{ABT3} and
at order $p^4$ the dispersive effect was about half of the correction for
$\eta\to3\pi$ makes it clear that a full order $p^6$
calculation was desirable.
The calculation \cite{BG2} generalized the methods
of \cite{ABT4} to deal with $\pi^0$-$\eta$ mixing.
The correction found in \cite{BG2} at order $p^6$ is 20-30\% in amplitude,
larger in magnitude than the dispersive estimates \cite{KWW,AL} but
with a shape similar to \cite{AL}.

The Dalitz plot in $\eta\to3\pi$ is parameterized in terms of $x$ and $y$
defined in terms of the kinetic energies of the pions $T_i$ and
$Q_\eta=m_\eta-2m_{\pi^+}-m_{\pi^0}$ for the charged decay and $z$ defined in
terms of the pion energies $E_i$. The amplitudes are expanded in
$x = \sqrt3 \left(T_+-T_-\right)/Q_\eta$, $y= 3T_0/Q_\eta-1$,
$z = (2/3)\sum_{i=1,3}\left(3 E_i-m_\eta\right)^2/\left(m_\eta-3m_{\pi^0}\right)^2$, via
\ba
|M(s,t,u)|^2 &=& A_0^2\left(1+ay+by^2+dx^2+fy^3+\cdots\right)\,,
%\nonumber\\
\quad
|\overline M(s,t,u)|^2 = \overline A_0^2
\left(1+2\alpha z+\cdots\right)\,.
\ea
Recent experimental results for these parameters are shown in 
Tabs.~\ref{tabDalitzcharged} and \ref{tabDalitzneutral}.
There are discrepancies among the experiments but the latest
precision
measurements of $\alpha$ agree.
\begin{table}
\begin{center}
\small
\begin{tabular}{|c|ccc|}
\hline
Exp. & a & b & d  \\
\hline
\rule{0cm}{12pt}KLOE  & $-1.090$\small$\pm0.005^{+0.008}_{-0.019}$ &
 $0.124$\small$\pm0.006\pm0.010$ & $0.057$\small$\pm0.006^{+0.007}_{-0.016}$\\
Crystal Barrel & $-1.22\pm0.07$ & $0.22\pm0.11$ &
$0.06$\small$\pm0.04$ (input) \\
Layter {\it et al.} & $-1.08\pm0.014$ & $0.034\pm0.027$ &
$0.046\pm0.031$ \\
Gormley {\it et al} &$-1.17\pm0.02$ & $0.21\pm0.03$ 
& $0.06\pm0.04$ \\
\hline
\end{tabular}
\end{center}
\caption{Measurements of the Dalitz plot distributions in 
$\eta\to\pi^+\pi^-\pi^0$. Quoted in the order cited in \cite{etacharged}. 
The KLOE result $f$ is $f=0.14\pm0.01\pm0.02$.}
\label{tabDalitzcharged}
\end{table}
\begin{table}
%\begin{center}
\centerline{
\small
\begin{tabular}{|c|ccccc|}
\hline
               & $A_0^2$ & a & b & d & f \\
\hline
LO             & 120 & $-1.039$ & $0.270$ & $0.000$   & $0.000$ \\
NLO             & 314 & $-1.371$  & $0.452$ & $0.053$ & $0.027$\\
NLO ($L_i^r=0$) & 235 & $-1.263$& $0.407$ & $0.050$ & $0.015$\\
NNLO           & 538 &   $-1.271$ & $0.394$ & $0.055$ & $0.025$ \\
%NNLOp          & 574 &   $-1.229$ & $0.366$ & $0.052$ & $0.023$ \\
%NNLOq          & 535 & $-1.257$ & $0.397$ & $0.076$ & $0.004$ \\
NNLO ($\mu=0.6$~GeV) & 543 & $-1.300$ & $0.415$ & $0.055$ & $0.024$\\         
NNLO ($\mu=0.9$~GeV) & 548 & $-1.241$ & $0.374$ & $0.054$ & $0.025$\\         
NNLO ($C_i^r = 0$)& 465 & $-1.297$  & $0.404 $ & $0.058$ & $0.032$ \\
NNLO ($L_i^r=C_i^r = 0$)& 251 & $-1.241$  & $0.424$ & $0.050$ & $0.007$ \\
\hline
\end{tabular}}
%\end{center}
\caption{Theoretical estimate of the Dalitz plot distributions in 
$\eta\to\pi^+\pi^-\pi^0$.}
\label{tabDalitzcharged_theory}
\end{table}
The predictions from ChPT to order $p^6$ with the input parameters
as described earlier are given in Tabs.~\ref{tabDalitzcharged_theory}
and \ref{tabDalitzneutral_theory}. 
%The predictions from the dispersive
%analysis as well as \cite{Borasoy} have not been included.
The different lines corresponds to variations on the input and the order of
ChPT. The lines labeled NNLO are the central results.
The agreement with experiment is not too good and clearly needs further study.
Especially puzzling is that $\alpha$ is consistently positive while
the dispersive calculations as well as \cite{Borasoy}
give a negative value.
The inequality $\alpha\le\left(d+b-a^2/4\right)/4$ derived in \cite{BG2}
shows that
$\alpha$ has rather large cancellations inherent in its prediction and that
the overestimate of $b$ is a likely cause of the wrong sign for
$\alpha$. The fairly large correction gives in the end
 larger values of $Q$ compared to those derived from the masses
\cite{BG2}.
\begin{table}
\begin{minipage}{0.56\textwidth}
\small
\begin{tabular}{|c|c|}
\hline
Exp. & $\alpha$\\
\hline
\rule{0cm}{12pt}
Crystal Ball (MAMI C) & $-0.032\pm0.003$\\
Crystal Ball (MAMI B) & $-0.032\pm0.002\pm0.002$\\
WASA/COSY  & $-0.027\pm0.008\pm0.005$\\
KLOE      & $-0.027\pm0.004^{+0.004}_{-0.006}$\\
%KLOE (prel) & $-0.014\pm 0.005\pm0.004$ \\
Crystal Ball (BNL)  & $-0.031\pm0.004$ \\
WASA/CELSIUS   & $-0.026\pm0.010\pm0.010$ \\
Crystal Barrel & $-0.052\pm0.017\pm0.010$ \\
GAMS2000  & $-0.022\pm0.023$ \\
SND  & $-0.010 \pm 0.021 \pm 0.010$\\
\hline
\end{tabular}
\normalsize
\caption{Measurements of the Dalitz plot distribution in
$\eta\to\pi^0\pi^0\pi^0$. Quoted in the order cited in \cite{etaneutral}.} 
\label{tabDalitzneutral}
\end{minipage}
\begin{minipage}{0.43\textwidth}
\small
\begin{tabular}{|c|cc|}
\hline
    & $\overline A_0^2$ & $\alpha$\\
\hline
 LO &   1090 &  $ 0.000 $ \\
 NLO &  2810 &  $0.013$\\
NLO ($L_i^r=0$) & 2100 &   $ 0.016 $ \\
 NNLO &  4790 &  $ 0.013$ \\
%NNLOq & 4790 & $ 0.014$ \\
NNLO ($C_i^r = 0$) & 4140 & $ 0.011$\\
NNLO ($L_i^r,C_i^r = 0$) & 2220 & $ 0.016$\\
\hline
dispersive \cite{KWW} & --- &\hskip-0.5cm $-(0.007$---$0.014)$\\
tree dispersive & --- & $-0.0065$\\
absolute dispersive& --- & $-0.007$\\
Borasoy \cite{Borasoy} & --- & $-0.031$\\
\hline
 error &  160 &  0.032 \\
\hline
\end{tabular}
\normalsize
\caption{Theoretical estimates of the Dalitz plot distribution in
$\eta\to\pi^0\pi^0\pi^0$. \cite{BG,BG2}}
\label{tabDalitzneutral_theory}
\end{minipage}
\end{table}

\section{Hard pion ChPT?}

In this section I discuss some recent work that argues that chiral
effects should also be calculable for processes with hard pions.
This type of arguments was given by Flynn and Sachrajda for $K_{\ell3}$
decays away from $q^2_{\textrm{max}}$ \cite{Flynn}. It was argued that those
arguments apply much more generally in \cite{BC} and there they were also
applied to $K\to\pi\pi$.

The underlying argument is that the main predictions of ChPT, namely
chiral logarithms come from soft pion lines. In pure ChPT as discussed
above the powercounting works since all lines are considered soft.
In the baryon sector, power counting has also been developed. There
the meson lines are soft and the baryons are always close to their mass-shell.
The heavy momentum always follows a baryon line. Here two momentum regions
are important, those close to the baryon momentum $p=M_Bv+k$ and the
soft ones $p=k$, $k$ soft for both.
As argued in \cite{GSS}, the $M_B$-dependence of
loop-diagrams is analytic and can be absorbed in the LECs.
This is a broad area of research as can be judged from many of the talks in
working group 2. Similarly, ChPT with mesons with a heavy-quark
relies on having two momentum regimes with one line always carrying the
large momentum, \cite{Wise,Donoghue}. Thereafter it has been argued that
ChPT could also be constructed for unstable particles near their
mass-shell \cite{JMW}, see especially the discussion in \cite{BGT}.
In all these cases, the underlying argument is always the same. The heavy-mass
dependence is analytic and can be absorbed in the LECs.

\begin{figure}
\begin{center}
\setlength{\unitlength}{0.8pt}
\begin{picture}(100,100)
\SetScale{0.8}
\SetWidth{3}
\Line(0,100)(20,80)
\Line(20,80)(20,20)
\Line(0,0)(20,20)
\Line(20,80)(80,80)
\Line(20,20)(80,20)
\Line(20,20)(80,80)
\Line(80,80)(100,100)
\Line(80,20)(100,0)
\SetWidth{1.}
\Curve{(20.,80.)(50.,95.)(80.,80.)}
\Line(80,80)(80,20)
\end{picture}
\raisebox{50\unitlength}{$\Rightarrow$}
\setlength{\unitlength}{0.8pt}
\begin{picture}(100,100)
\SetScale{0.8}
\SetWidth{3}
\Line(0,100)(20,80)
\Line(20,80)(20,20)
\Line(0,0)(20,20)
\Line(20,80)(80,80)
\Line(20,20)(80,20)
\Line(20,20)(80,80)
\Line(80,80)(100,100)
\Line(80,20)(100,0)
\SetWidth{1.}
\Line(20.,80.)(40,100)
\Line(60,100)(80,80)
\Line(80,80)(100,60)
\Line(100,40)(80,20)
%\Vertex(50,50){5}
\end{picture}
\raisebox{50\unitlength}{$\Rightarrow$}
\setlength{\unitlength}{0.8pt}
\begin{picture}(100,100)
\SetScale{0.8}
\SetWidth{3}
\Line(0,70)(50,50)
\Line(0,30)(50,50)
\Line(50,50)(100,70)
\Line(50,50)(100,30)
\SetWidth{1.}
\Line(50,50)(30,100)
\Line(50,50)(70,100)
\Line(50,50)(30,0)
\Line(50,50)(70,0)
\Vertex(50,50){8}
\end{picture}
\raisebox{50\unitlength}{$\Rightarrow$}
\setlength{\unitlength}{0.8pt}
\begin{picture}(100,100)
\SetScale{0.8}
\SetWidth{3}
\Line(0,70)(50,50)
\Line(0,30)(50,50)
\Line(50,50)(100,70)
\Line(50,50)(100,30)
\SetWidth{1.}
\Oval(50,75)(25,15)(0)
\Oval(50,25)(25,15)(0)
\Vertex(50,50){8}
\end{picture}
\end{center}
\caption{An example of the argument for a ``hard pion ChPT.'' 
The thick lines contain a
large momentum, the thin lines a soft momentum. 
Left: a general Feynman diagram with hard and soft lines.
Middle-left: we cut the soft lines to remove the soft singularity.
Middle-right: The contracted version where the hard part is
assumed to be correctly described
by a ``vertex'' of an effective Lagrangian.
Right:
the contracted version as a loop diagram. This is expected to reproduce
the chiral logarithm of the left diagram.}
\label{fig:power}
\end{figure}
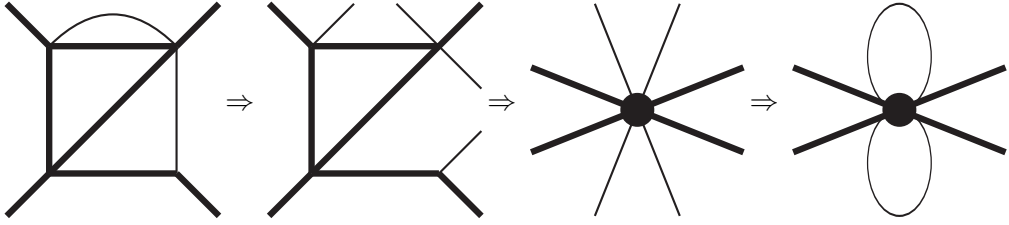

Ref.~\cite{BC} argued that the same type of reasoning
works for processes with hard pions. The reasoning 
is depicted in Fig.~\ref{fig:power}: Take a process with a given
external momentum configuration, identify the soft lines and cut them. The
resulting part is analytic in the soft stuff and should thus be describable
by an effective Lagrangian with coupling constants depending on the
given external momenta. The Lagrangian should be complete in the
\emph{neighbourhood}, in both momenta and processes, and should respect
the symmetries present in the problem. Loop diagrams with this effective
Lagrangian \emph{should} reproduce the nonanalyticities in the light masses.

In \cite{Flynn} and \cite{BC} it was proven that
respectively for  $K_{\ell3}$-formfactors and $K\to\pi\pi$-decays the
lowest order Lagrangian is sufficiently complete to be able to calculate
uniquely the pionic chiral logarithm. \cite{BC} explicitly kept some higher
order terms to illustrate the argument and found that
\ba
\label{resultNLO}
A_0^{NLO} &=& A_0^{LO}\left(1+\frac{3}{8F^2}\overline A(M^2)\right)+
\lambda_0 M^2+\mathcal{O}(M^4)\,,
\nonumber\\
A_2^{NLO} &=& A_2^{LO}\left(1+\frac{15}{8F^2}\overline A(M^2)\right)+
\lambda_2 M^2+\mathcal{O}(M^4)\,,
\ea
with $\overline A(M^2) = -1/(16\pi^2) M^2\log(M^2/\mu^2)$ and $M^2$
the lowest order pion mass. $\lambda_0$ and $\lambda_2$ depend on higher
order terms in the Lagrangian and are not calculable.
Another check was that the three-flavour ChPT did have the same pionic
chiral logarithms. Notice that the logarithms in (\ref{resultNLO})
are not due to the final state interaction, that effect goes into the
couplings in this approach, and actually go
against the $\Delta I=1/2$-rule.

\section{Conclusions}

ChPT in the meson sector is progressing and finds new application areas.
I have shortly reviewed two- and three-flavour ChPT for mesons
and discussed or provided a references to
several areas where there has been recent progress.

\acknowledgments

This work is supported in part by the European Commission RTN network,
Contract MRTN-CT-2006-035482  (FLAVIAnet), 
European Community-Research Infrastructure
Integrating Activity
``Study of Strongly Interacting Matter'' (HadronPhysics2, Grant Agreement
n. 227431)
and the Swedish Research Council. 
I thank the Berne group for organizing this very pleasant
meeting and so giving me the opportunity to visit Berne again.

\end{document}